\documentclass[prl,floatfix,twocolumn,showpacs,amsmath,amssymb]{revtex4}
\usepackage{graphicx}
\usepackage{dcolumn}
\usepackage{bm}
\begin{document}
\title{Ising transition driven by frustration in a 2D classical model with 
SU(2) symmetry}

\author{C\'edric Weber,$^{1,2}$ Luca Capriotti,$^{3}$ Gr\'egoire Misguich,$^{4}$ 
Federico Becca,$^{5}$ Maged Elhajal,$^{1}$ and Fr\'ed\'eric Mila$^{1}$}
\affiliation{
$^{1}$ Institut de Physique Th\'eorique, Universit\'e de Lausanne, 
CH-1015 Lausanne, Switzerland \\
$^{2}$ Institut Romand de Recherche num\'erique sur les Mat\'eriaux, 
Ecole Polytechnique F\'ed\'erale de Lausanne, CH-1015 Lausanne, Switzerland \\
${^3}$  Kavli Institute for Theoretical Physics, University of California, 
Santa Barbara CA 93106-4030 \\
${^4}$  Service de Physique Th\'eorique, CEA Saclay, 91191 Gif-sur-Yvette 
Cedex, France \\
${^5}$ INFM-Democritos, National Simulation Centre, and SISSA I-34014 
Trieste, Italy.
}

\date{\today}

\begin{abstract}
We study the thermal properties of the classical antiferromagnetic 
Heisenberg model with both nearest ($J_1$) and next-nearest ($J_2$) 
exchange couplings on the square lattice by
extensive Monte Carlo simulations. We show
that, for $J_2/J_1 > 1/2 $, thermal fluctuations give rise to an effective
$Z_2$ symmetry leading to a {\it finite-temperature} 
phase transition. We provide strong numerical evidence that this transition
is in the 2D Ising universality class, and that $T_c\rightarrow 0$ with
an infinite slope when $J_2/J_1\rightarrow 1/2$.
\end{abstract}


\maketitle

Since the milestone papers by Hohenberg and by Mermin and Wagner,~\cite{mermin} 
it is known that
in two-dimensional systems a continuous symmetry cannot be broken at 
any finite temperature, and only systems with a discrete symmetry can show a 
finite-temperature phase transition.
In this regard, a proposal by Chandra, Coleman and Larkin
(CCL)~\cite{chandra} opened a new route to finite-temperature phase transitions
in systems with a continuous spin-rotational invariance: CCL argued 
that the presence of frustrating interactions can induce non-trivial 
{\it discrete} degrees of freedom, that may undergo a phase transition at
low temperatures.
In particular, in Ref.~\onlinecite{chandra}, the authors considered the 
antiferromagnetic Heisenberg model with both nearest ($J_1$) and 
next-nearest neighbor ($J_2$) couplings:
\begin{equation} \label{hamilt}
\hat{\cal{H}}=J{_1}\sum_{n.n.}
\hat{{\bf {S}}}_{i} \cdot \hat{{\bf {S}}}_{j}
+ J{_2}\sum_{n.n.n.}
\hat{{\bf {S}}}_{i} \cdot \hat{{\bf {S}}}_{j}~,
\end{equation}
where $\hat{{\bf {S}}}_{i}$ are spin $S$ operators on a periodic square lattice 
with $N=L \times L$ sites. For $J_2/J_1 < 1/2$, the classical ground state has 
antiferromagnetic N\'eel order with pitch vector ${\bf Q}=(\pi,\pi)$, 
while for $J_2/J_1 > 1/2$, the classical 
ground state consists of two 
independent sublattices with antiferromagnetic order.~\cite{doucot} 
The ground state
energy does not depend on the relative 
orientations between the magnetizations of the two sublattices, and
the ground state has an SU(2)$\times$SU(2) symmetry.
Following Henley's analysis of the XY case,~\cite{henley} CCL
showed that both quantum and thermal fluctuations are expected
to lift this degeneracy by an {\it order by disorder} mechanism~\cite{villain}
and to select two collinear states which are the helical states with 
pitch vectors ${\bf Q}=(0,\pi)$ and $(\pi,0)$ respectively, reducing
the symmetry to SU(2)$\times Z_2$. CCL further argued that
the $Z_2$ symmetry should give rise to an Ising phase transition at finite
temperature and provided an estimate of the transition temperature.

The interest in this model raised recently with
the discovery of two vanadates which can be considered as 
prototypes of the $J_1{-}J_2$ model in the collinear region:
${\rm Li_2VOSiO_4}$ and ${\rm Li_2VOGeO_4}$.~\cite{millet,carretta}
Indeed, although the value of $J_2/J_1$ is not exactly 
known,~\cite{carretta,rosner,misguich} all estimates indicate that
$J_2 \gtrsim J_1$. In particular, NMR and muon spin rotation magnetization in 
${\rm Li_2VOSiO_4}$
provide clear evidence for the presence of a phase transition to a collinear order 
at $T_c \sim 2.8 K$. While several additional ingredients, like inter-layer
coupling and lattice 
distortion,~\cite{becca} are probably involved in the transition, the basic
explanation relies on the presence of
the Ising transition predicted by CCL.

However, CCL's predictions have been challenged by a number of numerical
studies. Using Monte Carlo simulations, Loison and Simon~\cite{loison} reported
the presence of two phase transitions for the $XY$ version of the classical model 
for $J_2/J_1 > 0.5$: 
A Kosterliz-Thouless transition, as expected for $XY$ models, followed by 
a transition which is continuous but does not seem to be
in the Ising universality class since their estimates of the 
critical exponents depend on 
the ratio $J_2/J_1$. More recently, the S=1/2 Heisenberg case has
been investigated by Singh and collaborators~\cite{singh}
using a combination of series expansion methods and linear spin-wave theory.
They  show that, if there is a phase transition, it can only occur at 
temperatures much 
lower than that predicted by CCL for S=1/2, and they argue that $T_c$ is
actually equal to zero.

In this Letter, we show, on the basis of extensive Monte Carlo simulations,
that the classical limit of the model of Eq.~(\ref{hamilt}), where spins
are classical vectors of length 1,  indeed
undergoes a continuous phase transition at a finite temperature,
that the critical exponents agree with the Ising universality 
class, and that, modulo minor adjustments of CCL's estimate,  
$T_c$ is in good {\it quantitative} agreement with CCL's prediction
in the range $J_1<J_2$ where their approximation is expected
to be valid. However, contrary to CCL's prediction, we show that $T_c$ goes 
continuously to zero when $J_2/J_1 \to 1/2$, and we argue that this is due to
a competition between N\'eel and collinear order at finite temperature in this
parameter range. 

Starting from the original spin variables $\hat{{\bf {S}}}_{i}$, we
construct the Ising-like variable of the dual lattice:
\begin{equation}\label{isingvar}
\sigma_\alpha = 
\frac{(\hat{{\bf {S}}}_{i} - \hat{{\bf {S}}}_{k}) \cdot
(\hat{{\bf {S}}}_{j} - \hat{{\bf {S}}}_{l})}
{\vert (\hat{{\bf {S}}}_{i} - \hat{{\bf {S}}}_{k}) \cdot
(\hat{{\bf {S}}}_{j} - \hat{{\bf {S}}}_{l}) \vert}~,
\end{equation}
where $(i,j,k,l)$ are the corners with diagonal $(i,k)$ and $(j,l)$ of the 
plaquette centered at the site $\alpha$ of the dual lattice.
The two collinear states with ${\bf Q}=(\pi,0)$ and ${\bf Q}=(0,\pi)$ have
$\sigma_\alpha= \pm 1$. 
It is important to stress that the normalization term does not affect the
critical properties of the model. It is only introduced to have a 
normalized variable. The Ising-like order parameter is defined as
$M_\sigma= (1/N)\sum_\alpha \sigma_\alpha$.

We have performed classical Monte Carlo simulations
using both local and global algorithms as well as more recent
broad histogram methods~\cite{brazilian} (details will be given 
elsewhere~\cite{cedric}) to calculate the temperature and size dependence
of several quantities including the Binder cumulant, the susceptibility
and the correlation length associated to $M_\sigma$, as well as the 
specific heat, for sizes up to $200 \times 200$ and for several values 
of $J_2/J_1$ between 1/2 and 2. For reasons discussed below the critical 
behaviour is easier to detect for small values of $J_2/J_1$, and we 
first concentrate on $J_2/J_1=0.55$.

\begin{figure}
\vspace{-0.3cm}
\includegraphics[width=0.4\textwidth]{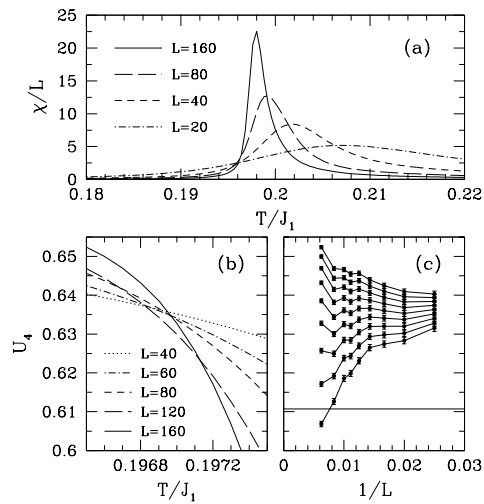}
\vspace{-0.3cm}
\caption{\label{fig:divergence}
(a) Finite-size susceptibility $\chi(L)/L$ and (b) Binder cumulant $U_4$ 
as a function of the temperature for 
different sizes; (c) Binder cumulant $U_4$ as a function of $1/L$ for different
temperatures (0.1965,0.1966, \dots , 0.1973 from top to bottom). Lines are guides
to the eye and the horizontal line marks the value for $U_4$ at
the critical point for the Ising universality 
class.~\protect\cite{u4univ} All data are for $J_2/J_1=0.55$.
}
\end{figure}

As a first hint of a phase transition, we report the temperature dependence
of the susceptibility defined by
$\chi=(N/T) (\langle M_\sigma^2 \rangle - \langle |M_\sigma| \rangle^2)$
for different sizes [see Fig.~\ref{fig:divergence}(a)].
If there is a phase transition, this susceptibility is expected to diverge 
at $T_c$ in the thermodynamic limit, and indeed, the development
of a peak around $T/J_1=0.2$ upon increasing the system size is
clearly visible.
To get a precise estimate of $T_c$, we have calculated Binder's fourth 
cumulant of the order parameter defined by:
$U_4(T) = 1 - \langle M_\sigma^4 \rangle / 3 \langle M_\sigma^2 \rangle^2$.
This cumulant should go to 2/3 below $T_c$ and to zero above $T_c$ when 
the size increases, and the finite-size estimates are expected to cross 
around $T_c$. Binder cumulants for different sizes are reported in 
Fig.~\ref{fig:divergence}(b), and they indeed cross around $T/J_1\simeq 0.197$. 
In Fig.~\ref{fig:divergence}(c), we report $U_4(T)$ as a function of $1/L$ 
for several temperatures around $0.197$. Excluding temperatures for which
$U_4$ clearly increases or decreases with $1/L$ leads to the remarkably 
precise estimate $T_c/J_1=0.1970(2)$.

\begin{figure}
\vspace{-2.2cm}
\includegraphics[width=0.4\textwidth]{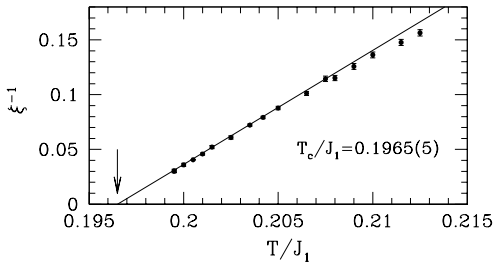}
\vspace{-1.0cm}
\caption{\label{fig:xivstemp}
Inverse of the correlation length $\xi^{-1}$ as a function of the temperature 
for different sizes of the lattice and $J_2/J_1=0.55$. 
The critical exponent $\nu$ and $T_c$ can be extracted from the behavior 
of $\xi^{-1}$ as a function of the temperature. The arrow marks the
resulting $T_c$.
}
\end{figure}

To identify the universality class of the phase transition, we have looked
at the finite-size scaling of several quantities. The critical exponents $\nu$ 
and $\gamma$ can be extracted from the dependence
of the peak position of the susceptibility 
$T_c(L) = T_c + a \times L^{-1/\nu}$ and from its value
$\chi(L,T_c) \sim L^{\gamma/\nu}$ as a function of $L$. Using the value of
$T_c$ deduced from Binder's cumulant, the fits lead to 
$\nu=1.0(1)$ and $\gamma/\nu = 1.76(2)$. 
A more precise estimate of the exponent $\nu$ can be obtained
from the temperature dependence of the second-moment correlation 
length $\xi$~\cite{olsson}, extracted from the Fourier components of the 
correlation functions of the Ising-like variable (\ref{isingvar}).
By considering only values such that 
$\xi \lesssim L/6$, where the finite-size effects are found to be negligible,
it is possible to have a very accurate
value of the critical exponent from the fact that, for $T \gtrsim T_c$,
$\xi^{-1} = A (T-T_c)^\nu$.
In Fig.~\ref{fig:xivstemp}, we report the behavior of the correlation
length $\xi$ as a function of the temperature. By performing a 
three-parameter fit (for $A$, $T_c$ and $\nu$) we obtain $T_c/J_1=0.1965(5)$ 
and $\nu=1.00(3)$.
This value of $T_c$ is compatible with the
estimation given by Binder's cumulant.

\begin{figure}
\vspace{-0.3cm}
\includegraphics[width=0.4\textwidth]{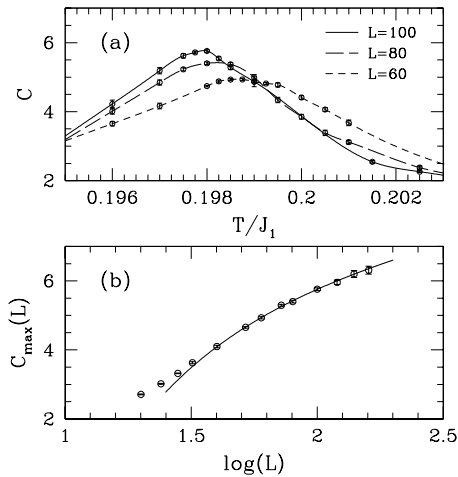}
\vspace{-0.3cm}
\caption{\label{fig:specheat}
(a) Specific heat per site $C$ as a function of the temperature for
different sizes of the lattice and $J_2/J_1=0.55$. The lines are giudes to the
eye. (b) Maximum of the specific heat per site $C_{{\rm max}}(L)$ as a function 
of $L$. The line is a three-parameter fit (see text).
}
\end{figure}

These exponents agree with those of the Ising universality 
class in 2D ($\nu=1$ and
$\gamma=7/4$). 
A cross-check for this universality class comes from the measure 
of the critical exponent $\alpha$, related to the
divergence of the specific heat per site, 
$C_{{\rm max}}(L) \sim L^{\alpha/\nu}$.
Indeed, the value of $\alpha$ is the fingerprint for the 2D Ising universality 
class, for which we have $\alpha=0$ and a logarithmic divergence of the 
specific heat as a function of $L$.
In Fig.~\ref{fig:specheat}, we show the results for the specific heat per site.
We have 
obtained a very accurate fit of the maximum of the specific heat per site 
as a function of $L$ with the known expression for
the leading finite-size corrections of the 2D Ising model,~\cite{ferdinand}
$C_{{\rm max}}(L)=a_0+a_1 \log(L) + a_2/L$, 
with $a_0$, $a_1$ and $a_2$ fitting parameters, while
a power law is clearly inadequate. These results are consistent with $\alpha=0$.

Finally, if the phase transition is indeed Ising, Binder's cumulant at $T_c$
should reach the universal value $U_4(T_c) \sim 0.6107$ in the thermodynamic 
limit.~\cite{u4univ} 
The non-montonous behaviour of Binder's cumulant with the size 
prevents a precise extrapolation, but this
value is not incompatible with our numerical data 
[see Fig.~\ref{fig:divergence}(c)]. 

\begin{figure}
\vspace{-2.2cm}
\includegraphics[width=0.4\textwidth]{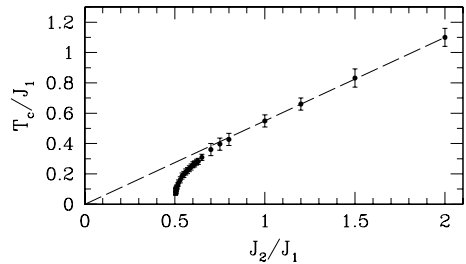}
\vspace{-1.0cm}
\caption{\label{fig:phasediag}
Monte Carlo results for the critical temperature $T_c$ as a function 
of the frustrating ratio $J_2/J_1$. The line is an extrapolation of
the large $J_2$ data down to $J_2=0$ (see text).
}
\end{figure}

Altogether, these results show unambiguously that a phase transition
occurs for $J_2/J_1=0.55$ at $T_c=0.1970(2)$ and give strong evidence in
favour of 2D Ising universality class, in agreement with CCL's prediction. 
The same analysis can be repeated for different values of the 
frustrating ratio $J_2/J_1$, and the complete phase diagram is shown in
Fig.~\ref{fig:phasediag}, where we report $T_c$ as a function of $J_2/J_1$.
While the critical behaviour is everywhere consistent with Ising, it turns
out that finite-size effects become more and more severe upon increasing
$J_2/J_1$, preventing a meaningful determination of $T_c$ with available
cluster sizes beyond $J_2/J_1\simeq 2$. Indeed, for large ratios
$J_2/J_1$, physical quantities 
such as the susceptibility and the specific heat only exhibit broad
peaks  while the mean value of the order parameter goes very smoothly to 
zero, a behaviour typical of strong finite-size effects. This we believe 
can be traced back
to the width of the domain walls between domains with 
${\bf Q}=(\pi,0)$ and ${\bf Q}=(0,\pi)$,
which we have estimated by studying systems with fixed boundary conditions.
Details will be presented elsewhere,~\cite{cedric}
but the width increases very fast with $J_2/J_1$, starting around 10
lattice spacings
for $J_2/J_1 \gtrsim 1/2$, and already reaching values
of the order of 40 lattice spacings for $J_2/J_1 \simeq 1.5$. Since cluster
sizes should be significantly larger than the width of the domain
walls to observe the critical behaviour, we are not able to go 
beyond $J_2/J_1\simeq 2$.

Let us now discuss the dependence of $T_c$ upon $J_2/J_1$ 
(see Fig.~\ref{fig:phasediag}). Two regimes can clearly be
identified: (i) A large $J_2$ regime ($J_2/J_1>1$) where $T_c/J_1$ scales 
more or less linearly with $J_2/J_1$; (ii) A smaller $J_2$ regime defined by
$(J_2/J_1-1/2)\ll 1$
where $T_c$ vanishes with an infinite slope as $J_2/J_1 \to 1/2$. 
For $J_2/J_1 \sim 1/2$, this
disagrees with CCL's prediction that $T_c/J_2$ reaches its 
maximum when $J_2/J_1 \to 1/2$. However, since their approach 
is based on an expansion in $J_1/J_2$, this is not a final blow, and
CCL's predictions should be tested in the large $J_2$ regime, where
the approximations are better controlled. CCL's central criterion
for estimating $T_c$ is the equation $T_c \simeq E(T_c) (\xi_N(T_c)/a)^2$,
where $E(T_c)$ is the energy barrier to go from one domain to the other
through the intermediate canted state where sublattice staggered magnetizations
are perpendicular, and $\xi_N(T)$ is the N\'eel correlation length of each 
sublattice.
To get a quantitative estimate of $T_c$ based on this approach, we 
have solved this equation using the exactly known temperature dependence 
of $\xi_N(T)$ for the classical antiferromagnet~\cite{caracciolo} and a 
corrected expression for $E(T)$.~\cite{note} This leads to 
$T_c = 0.768\, J_2 / (1 + 0.135\, \ln (J_2/J_1))$. The best way to check this 
approach would be to detect the logarithmic correction, but unfortunately
this would require to go to temperatures much larger than what we can 
reach, and in the temperature range available to our simulations, CCL's
prediction reduces to $T_c\simeq 0.77\, J_2$. This prediction is in good agreement
with our results: In the large $J_2$ regime, $T_c$ indeed scales linearly with 
$J_2$, within error bars our high $J_2$ data extrapolate to 0 at $J_2=0$,
and the slope is equal to 0.55. Note that the slight
difference in slopes is not significant since including 
a constant factor in front of $E(T_c)$ in the self-consistent equation for 
$T_c$ would change the slope. So altogether we believe that the
present results support CCL's prediction when  $J_2/J_1$ is not too close to 
1/2. In that respect, we note that a similar estimate of $T_c$ can be performed
on the basis of CCL's criterion for S=1/2 using recent estimates of 
$\xi_N(T)$ for the S=1/2 Heisenberg antiferromagnet on the square 
lattice,~\cite{kim}
which leads to $T_c=0.496\, J_2/(1+0.78\, \ln(J_2/J_1))$. These estimates are 
considerably lower than those used in Ref.~\onlinecite{singh}, and with these
new estimates we believe that a phase transition cannot be excluded on the basis
of the numerical results of Ref.~\onlinecite{singh}. In fact, although the
precise values of $J_1$ and $J_2$ in LiVOSiO$_4$ are still a matter of 
discussion,~\cite{carretta,rosner,misguich} we note that the order of magnitude
of these new estimates is consistent with the experimental results of
Ref.~\onlinecite{carretta}.
 
Finally, let us discuss the behaviour of $T_c$ close to $J_2/J_1=1/2$. 
The disagreement with CCL suggests that one ingredient is missing
in that range. In fact, we found that the fluctuations 
have a clear N\'eel character when $J_2/J_1 \sim 1/2$, 
implying that, although the collinear phase is energetically
favoured at zero temperature when $J_2/J_1>1/2$, thermal fluctuations
favour the N\'eel state, as quantum
fluctuations indeed do.~\cite{mila} We then expect a crossover to take place
between a high-temperature N\'eel phase and a low temperature collinear phase.
To estimate the crossover temperature, we start from the low-temperature 
expansion of the free energy per site for classical spins which reads,
when only quadratic modes are thermodynamically relevant,
\[
f=e_0- T \ln T - \left(\frac{1}{N} \sum_q \ln \omega_q\right)\, T 
+ a_2 T^2 + \dots,
\]
where $e_0$ is the ground-state energy per site  and $\omega_q$ are the frequencies of 
the quadratic modes. 
For the present purpose, the coefficients of this expansion 
should be determined in the 
limit $J_2/J_1\rightarrow 1/2$ from below and above for the N\'eel and
collinear states respectively, and the cross-over temperature $T_0$ is
determined by $f_{\rm Neel}=f_{\rm col}$. It turns out that,
in the limit $J_2/J_1\rightarrow 1/2$,
$\sum_q \ln \omega_q$ has the same value for the N\'eel and collinear phases, 
so that the linear term drops from this equation, leading to
$T_0\propto (J_2/J_1-1/2)^{1/2}$ since $e_0^{\rm Neel}=-2J_1+2J_2$ and 
$e_0^{\rm coll}=-2J_2$.[\onlinecite{note2}]
Now, the Ising transition can only occur below $T_0$ since the system should
already have short-range collinear fluctuations. Then, when the intrinsic 
temperature scale of the Ising transition as determined by CCL is larger 
than $T_0$, we expect the 
transition temperature to be of the order of $T_0$. This argument thus
predicts that, as $J_2/J_1\rightarrow 1/2$, $T_c$ should vanish with an 
infinite slope and
with an exponent equal to 1/2.[\onlinecite{note2}] 
This is in qualitative agreement with our numerical data, which clearly indicate
that the slope is infinite, and are consistent with 
an exponent in the range $0.3-0.5$.
Large scale numerical simulations are in progress to try to refine this estimate.

In conclusion, we have established the presence of a finite-temperature phase
transition in the classical $J_1-J_2$ Heisenberg model on a square lattice when
$J_2/J_1>1/2$, we have provided strong arguments in favour of Ising universality
class, and we have determined $T_c$ as a function of $J_2/J_1$ with high
accuracy, showing in particular that it vanishes with an infinite slope
when $J_2/J_1\rightarrow 1/2$. These results, together with revised estimates
of $T_c$ in the S=1/2 case following CCL, are expected to set the stage 
for further theoretical and experimental investigations. 

We acknowledge useful discussions with S. Sachdev and M. Zhitomirsky, 
and we are grateful to 
P. Coleman for giving us some information on unpublished material
related to Ref.~\onlinecite{chandra}. This work was supported by NSF under
Grant No.DMR02-11166 (L.C.), INFM (F.B.),  MaNEP (M.E.) and by the Swiss National 
Fund (C.W. and F.M.). G.M. acknowledges the hospitality of IRRMA.


\begin{thebibliography}{99}
\bibitem{note} According to our derivation, the values given by CCL for
their coefficients $\gamma_Q$ and $\gamma_T$ should be divided by 4.
\bibitem{note2} This argument is based on the assumption that 
$a_2^{\rm col}\ne a_2^{\rm Neel}$. If they are equal, the
exponent will be smaller.
\end{thebibliography}
\end{document}